\documentclass[conference]{IEEEtran}
\usepackage{graphicx,latexsym}
\ifCLASSINFOpdf
\else
\fi
\begin{document}
%
\title{Cellulose-Bound Magnesium Diboride Superconductivity}

\author{\IEEEauthorblockN{Y.L. Lin and M.O. Pekguleryuz}
\IEEEauthorblockA{Department of Mining \& Materials Engineering\\
McGill University\\
Montreal, QC Canada
}
\and
\IEEEauthorblockN{J. Lefebvre, C.J. Voyer, D.H. Ryan and M. Hilke}
\IEEEauthorblockA{Department of Physics\\Center for the Physics of Materials\\
McGill University\\
Montreal, QC Canada\\
}}


%


\maketitle

\begin{abstract}
Two-phase superconductor tapes were produced by blending high purity magnesium diboride powder with a liquid ethylcellulose-based polymeric binder. This procedure produced a material which is easily formable with a high superconducting transition temperature (38K). We show that the bulk superconducting properties are not affected by the presence of the binder, nor is there any evidence of a chemical reaction between the superconducting particles and the binder. However, the transport properties of the material are strongly affected by the presence of the binder, which leads to a seven order of magnitude increase of the normal state resistance along with a seven order of magnitude decrease of the transport critical current density. This new material is shown to be equivalent to a system of coupled Josephson junctions.

\end{abstract}


%
\IEEEpeerreviewmaketitle

\section{Introduction}
Magnesium diboride was found to be superconducting in 2001 by Nagamatsu et al. \cite{1}. In crystalline or polycrystalline form MgB$_2$ has a transition temperature of 39 K \cite{1}, the highest for any simple intermetallic compound in addition of being one of the lightest. It boasts a high critical current density, surpassing 85,000 A/cm2 \cite{2}, a high critical magnetic field, typically 8 T at 21 K \cite{3} and grain boundary transparency \cite{4}. This leads to a huge potential for applications \cite{magnet} in view of replacing other very successful conventional superconductors based on niobium alloys, which all have lower critical temperatures but have been used as workhorses for superconducting applications like magnets. It turns out that MgB$_2$ is a simple compound, which was already discovered in 1953 \cite{5}. The wide availability of this compound eliminates many steps in its fabrication, including sintering and pressing,  synthesis at elevated temperatures and under reducing conditions, as is the case for other high temperature cuprate superconductors. The pairing mechanism is meditated by phonons, which makes it a conventional BCS superconductor \cite{BCS} with the exception of it's two double gap structure \cite{double}. Thanks to its conventional pairing mechanism it would be an ideal candidate for the next generation of quantum interferences devices based on isotropic tunneling in Josephson junction.

A major roadblock in the development of magnesium boride in its quest to supplant other conventional compounds is the relative difficulty in producing reliable superconducting wires. While some techniques such as the powder in tube method \cite{PIT} have proven to work, they remain very brittle, similar to high temperature cuprate superconductors. The idea in this work is to explore new avenues in order to fabricate a more robust and malleable material, with a good potential for quantum interference devices and which can be easily molded in any shape. Therefore it is important to have an isotropic material, where the properties do not depend on particular orientations. Hence, we explored the possibility of obtaining such a material by using readily available polycrystalline MgB$_2$ powder as starting material and bind it using a cellulose based polymer. Cellulose tends to react only very weekly and is very robust, hence constitutes a good candidates for forming novel compounds.

\section{Fabrication}

\begin{figure}[!t]
\centering
\includegraphics[width=3in]{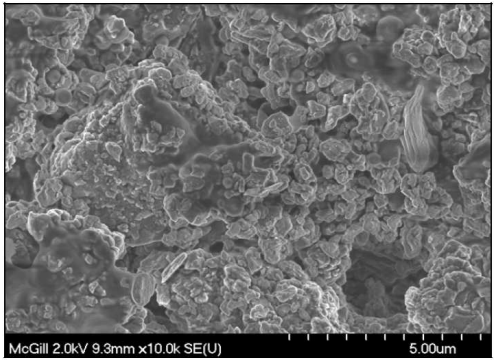}
\caption{SEM image of the MgB$_2$ polymer tape showing surface topography and MgB2 particles}
\label{SEM1}
\end{figure}

The MgB$_2$ powder was incorporated into a flexible polymeric matrix, ethylcellulose which can be naturally derived and shaped into any geometry. The binder was prepared from a mixture of Standard 45 Ethocel and a 50/50 solvent blend of butyl Cellosolve and terpineol. The MgB$_2$ powder was commercially obtained and of 99.8\% purity, of particles below 44 microns and was incorporated with the prepared wet binder in ratios varying from 20\% to 90\%. In order to produce tapes, the MgB$_2$ paste was spread onto an Al foil substrate and then dried before depositing contacts. Some samples were also deposited directly on a plastic sample holder, which did not affect the quality of the samples. Shaping the tapes required a simple cutting tool and peeling from the flexible aluminium substrate. The surface of the tapes was imaged at high magnification using a Hitachi 4700-S field emission gun scanning electron microscope (FEGSEM) shown in figure \ref{SEM1}. The surface is rough due to the polymer component as well as the jaggedness of the MgB$_2$ particles. This method is quite different from the powder in tube method since no inter-diffusion is occurring.

\begin{figure}[!t]
\centering
\includegraphics[width=3in]{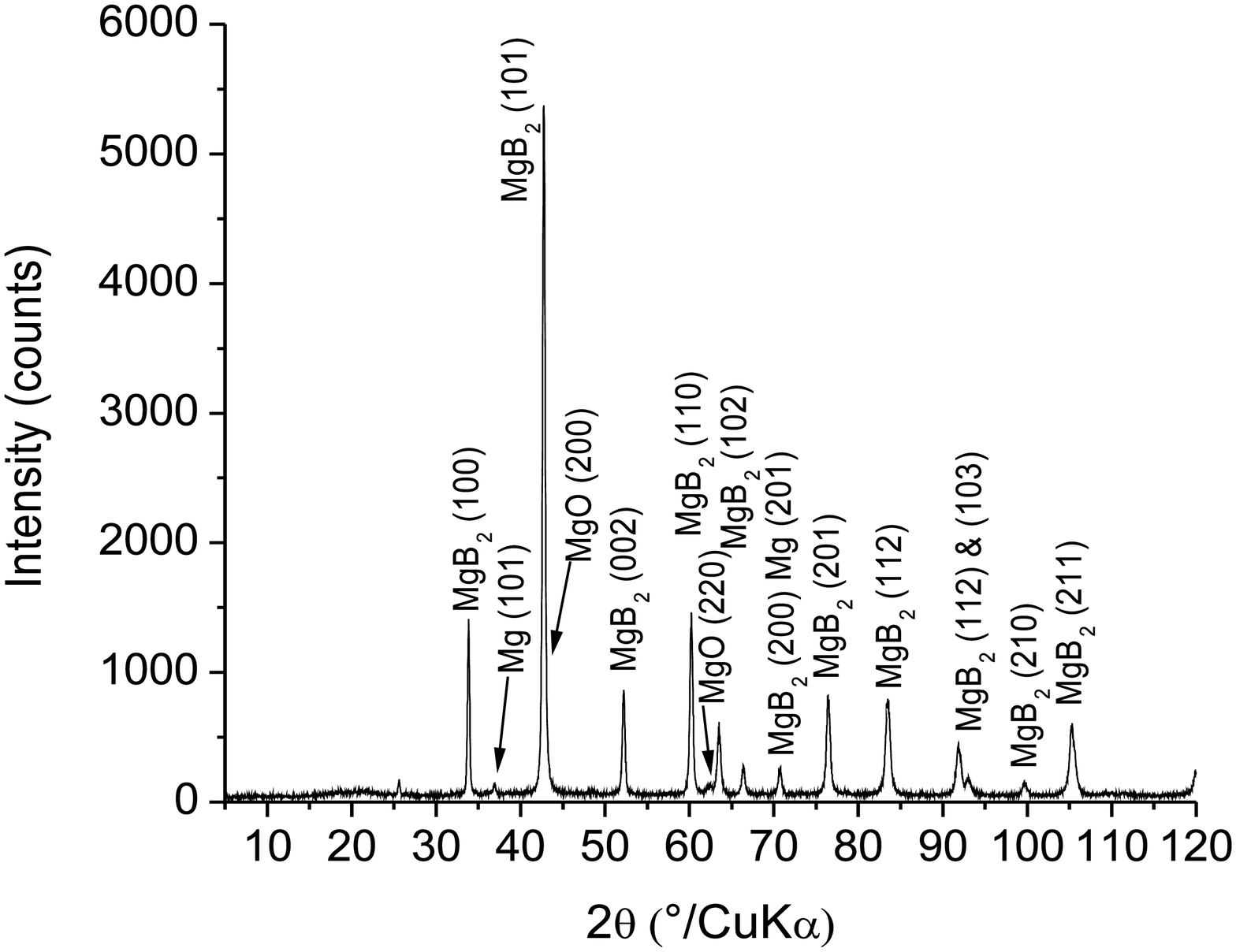}
\caption{X-ray diffraction spectrum of a MgB$_2$ polymer tape. }
\label{XRD}
\end{figure}

The elemental and compound identifications were determined using energy dispersive spectroscopy (EDS) and x-ray diffraction (XRD). XRD was performed using a Philips 1050/65 x-ray diffractometer and a Philips PW 1710 diffractometer. The American Instrument Inc. x-ray generator was the source of CuK$\alpha$ radiation. The spectrum is shown in figure \ref{XRD}. Except for some additional traces of magnesium oxide, no other elements other than MgB$_2$ are seen. More importantly, there is no change in the spectrum when comparing X-ray diffraction of the powder directly with the diffraction of the MgB$_2$ polymer tape, indicating that no structural change is occurring in the polymer mix.

The elemental analysis, using EDS showed expected peaks for B, C, O and Mg. The carbon and oxygen peaks are attributed to the polymer component of the sample, whose constituent compounds are ethylcellulose, [C$_6$H$_7$O$_2$(OC$_2$H$_5$)$_3]$n, and butyl acetic Cellosolve C$_4$H$_9$OCH$_2$CH$_2$OC(O)CH$_3$.

\begin{figure}[!t]
\centering
\includegraphics[width=3in]{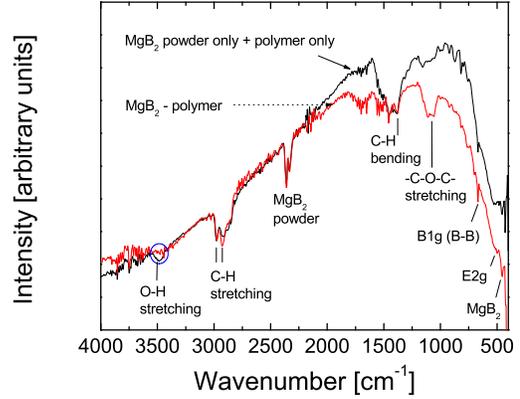}
\caption{FT-IR intensity for the MgB$_2$-polymer tape (red) and the sum of the intensities for the MgB$_2$ powder only plus the polymer only.}
\label{FTIR}
\end{figure}

In order to investigate the possibility of a chemical change in the ethylcellulose polymer during fabrication, Fourier transform infrared spectroscopy (FT-IR) analysis was conducted shown in figure \ref{FTIR}. The dried binder, the MgB2 powder and the MgB2 polymer tapes were analyzed via FT-IR separately and the strongest peaks were located and compared to the MgB$_2$ polymer tapes. The FT-IR instrument was a Bio-Rad FTS 6000 spectrometer, used at 2.5 kHz with a resolution of 8 cm$^{-1}$ for 128 scans in the range of 4000 to 450 cm$^{-1}$. Carbon black was used to determine the background spectrum. The signal was amplified using a photoacoustic (PA) accessory. In addition, the MgB$_2$ polymer tapes were crushed for greater signal amplification before placing in the PA cell. We found peaks at (1056, 1383, 2844, 2930, 3482) cm$^{-1}$, when measuring only the polymer, which is consistent with the results by Suthar et al. \cite{6} for ethylcellulose. The same peaks were found again in the FT-IR spectrum of the MgB$_2$-polymer tapes, showing that there is no chemical change of the ethylcellulose when mixed to the MgB$_2$ powder. The only change was a substantial weakening of the peak at 3482cm$^{-1}$, which corresponds to the O-H stretching as indicated by the circle on figure \ref{FTIR}. The MgB$_2$-polymer tapes showed additional peaks at (506, 671, 1459, 1650, 671) cm$^{-1}$, corresponding to the MgB$_2$ system. We found these same peaks when measuring only the MgB$_2$ powder, which demonstrates that the ethylcellulose did not affect the MgB$_2$ system either. The positions of the peaks were all consistent with the ones measured by other groups in MgB$_2$ \cite{7,8,9}. Surprisingly, no new peak was found between 450 and 4000 cm$^{-1}$, which might correspond to a new bond created between the polymer and MgB$_2$. Overall, the FT-IR measurements showed that mixing MgB$_2$ powder with ethylcellulose does not alter its components significantly.

\section{Bulk superconductivity}

\begin{figure}[!t]
\centering
\includegraphics[width=3in]{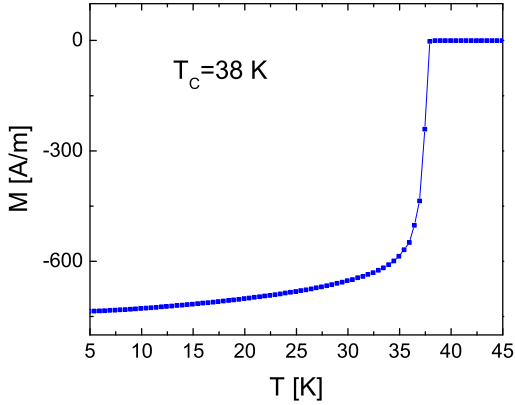}
\caption{The Magnetization at low fields for a MgB$_2$-polymer tape as a function of temperature.}
\label{M}
\end{figure}

Bulk superconductivity was tested by performing standard magnetization measurements as a function of temperature and field. The temperature dependence shows a sharp drop at $T_C=38\pm0.5$K as shown in figure \ref{M}  and indicates high quality superconducting MgB$_2$ particles. The dependence is very similar to observations of single polycrystals of MgB$_2$ \cite{Shigeta}. This means that the cellulose based binder does not alter the superconducting properties of the MgB$_2$ particles.

\begin{figure}[!t]
\centering
\includegraphics[width=3in]{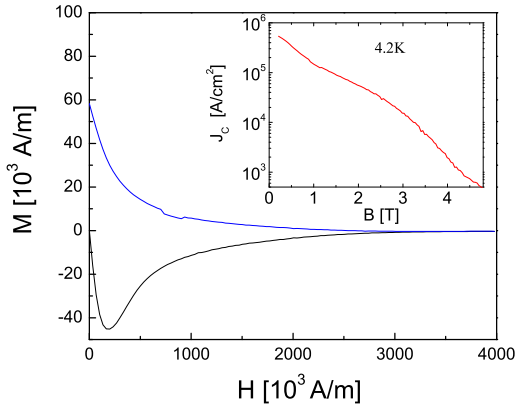}
\caption{The hysteresic magnetization dependence as a function of applied magnetic field. The lower curve corresponds to the up-sweep, whereas the top curve shows the down sweep.}
\label{Jc}
\end{figure}

For magnetic fields beyond the first critical field $Bc_1$, it is possible to extract the bulk critical current density from the hysteric magnetization curve using Bean's model and assuming that the polycrystals in the binder are well separated. For cylindrical-like grains the critical current density is then given by $J_C=3\Delta M/a$ \cite{Bean}, where the factor 3 is a geometrical factor assuming a cylindrical shape. This geometrical factor varies by less than a factor of two for any standard geometry with a similar aspect ratio. Hence, the order of magnitude of the extracted current is insensitive to the details of our assumptions. $\Delta M$ is the difference in magnetization in international units [A/m] of the up-sweep and the down-sweep. $a$ (in meter) is the average diameter of the particle in the direction perpendicular to the applied field. $J_C$ shown in figure \ref{Jc}, therefore represents the magnetization critical current density of a single superconducting particle. It is also possible to extract the susceptibility of the entire sample, by evaluating the slope of the initial low field magnetization curve, which we found to be $\chi=-0.45\pm0.05$. In order to obtain this value we used an effective demagnetizing factor corresponding to a vertical cylinder of aspect ratio 1 \cite{Brandt}. This implies that only about half (by volume) of the sample is superconducting, i.e., the binder does not contribute to superconductivity. Indeed, the density of the MgB$_2$ polymer was (1.7$\pm$0.2)g/cm$^3$ as opposed to 2.6g/cm$^3$ for the density of a MgB$_2$ crystal, which amounts to 60\% of MgB$_2$ particles and is comparable to the measured susceptibility.

Assuming a typical particle diameter of 40$\mu m$ we obtain a maximum critical current density of $J_c\simeq 10^6$A/cm$^2$, which is very similar to the results for a single polycrystal \cite{Shigeta} of a similar size. This shows that the superconductivity within a particle is not affected by the presence of the cellulose based binder. This is however in stark contrast, as we will see below, to the transport properties, where the coupling between superconducting particles becomes important.

\section{Transport superconductivity}

\begin{figure}[!t]
\centering
\includegraphics[width=3in]{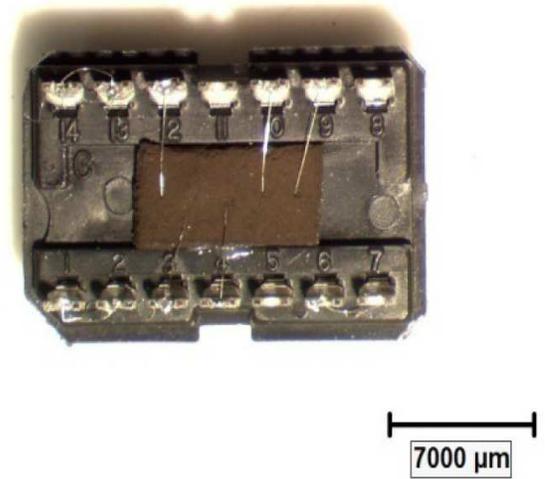}
\caption{Image of the MgB$_2$ polymer with contacts on a probe header.}
\label{SamplePic}
\end{figure}

In order to measure the transport properties we made ohmic contacts using two different methods. The simplest consists in placing silver wires in the wet MgB$_2$ powder-binder mix and letting it dry and solidify. Such a sample is shown in figure \ref{SamplePic}, where it is mounted on a 14pin header for transport measurements. Alternatively, silver epoxy was used as well.

\begin{figure}[!t]
\centering
\includegraphics[width=3in]{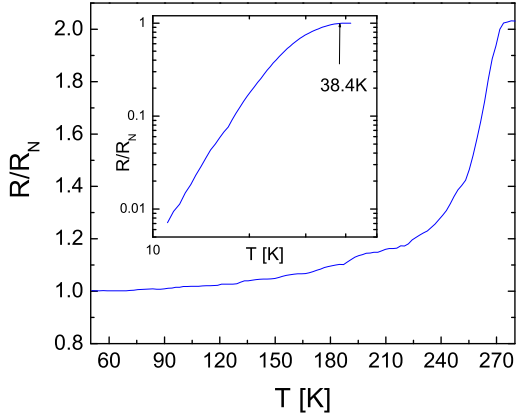}
\caption{Resistance normalized to the resistance just above $T_C$ as a function of temperature. The main figure shows the dependence of the high temperature part, whereas the inset shows the temperature dependence of the normalized resistance on a logarithmic scale close to the superconducting transition. }
\label{RvsT}
\end{figure}

Quite strikingly, the temperature dependence of the resistance is very weak above the critical temperature, before dropping very slowly into the superconducting regime as shown in figure \ref{RvsT}. While the onset is at $T_C=38.4\pm 0.5$K and very close to the value obtained from magnetization measurements, the transition width as measured from the 10\% and 90\% value of the normal resistance is $\Delta T=15\pm 3$K. The errors are based on the statistics of a dozen tested samples all showing very wide transitions. This has to be contrasted to the case of polycrystals or single crystals, where the temperature drop is very sharp.

Similarly striking is the extremely high value of resistivity above $T_C$. Indeed, our samples show a range in resistivities of $\rho=3-20\Omega$ cm, which is about 7 orders of magnitude higher than typical values for crystals and polycrystals ($\rho\simeq 2.5\mu \Omega$cm) \cite{Xu}. A typical resistance for our sample geometry is 100$\Omega$. Moreover, the temperature dependence of the normal resistance is very weak just above $T_C$. It varies by less than 20\% up to 200K. This again is very different to single crystals and polycrystals, where the residual resistance ratio (RRR) factor is about 5 and 20, respectively. Above 200K, some of our samples show almost no temperature dependence (less than 20\%), whereas other samples show an increase up to a factor 10. We believe that is probably due to the mechanical rearranging of some particles.

When comparing the transport critical current density, as determined from the current voltage characteristics, to the critical current density extracted from the magnetization data we obtain a transport critical current density in the range $J_C$=0.05-0.2 A/cm$^2$, which is 7 orders of magnitude smaller than the magnetization critical current density.

Hence, this leaves us to draw the following picture of what is happening in this superconducting compound: The MgB$_2$ superconducting particles are largely unaffected by the mixing with the cellulose based binder, which holds the superconducting particles together. Macroscopic superconductivity is obtained thanks to Josephson coupling between these particles. This coupling is strong enough to allow for macroscopic superconductivity to occur. The presence of the Josephson junctions (JJ) in the system explains the wide transition observed around the critical temperature.

\begin{figure}[!t]
\centering
\includegraphics[width=3in]{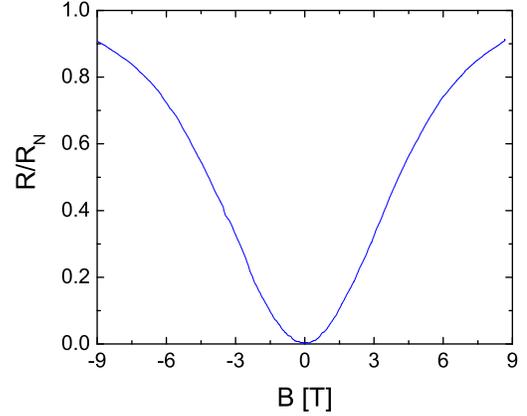}
\caption{The normalized resistance as a function of the magnetic field at 4.2K}
\label{RvsB}
\end{figure}

Further evidence that this system is dominated by a network of JJ is the nature of the magnetic field induced transition. As shown in figure \ref{RvsB}, the magnetic field induced transition from normal to superconducting is extremely wide (about 6T) as measured from the 10\% to 90\% value of the normal resistance. This is very different from single crystals or polycrystals, which exhibit a much narrower transition range.

This picture of a network of JJ also explains the very low values of critical transport currents observed. Indeed, the Ambegaokar Baratoff relation \cite{JJ} for JJ implies an inverse dependence between the critical current and the normal resistance ($J_C\sim R_N^{-1}$). Hence, the 7 order of magnitude increase of the normal resistance as compared to the single crystal can be directly related to the observed 7 order magnitude decrease in critical currents, providing further evidence that our sample structure is dominated by JJ. Interestingly, varying the ratio of the cellulose binder to MgB$_2$ does not significantly alter the transport properties, indicating that the coupling between superconducting particles is not affected by the ratio of cellulose to MgB$_2$. Therefore, the cellulose binder simply "holds" the superconducting particles together without modifying any of the intrinsic properties. Within this picture it is then possible to greatly enhance the critical current density by simply reducing the size of the superconducting particles, which would lead to an increase proportional to $J_C\sim a^{-2}$ \cite{JJN}, where $a$ is the typical diameter of the superconducting particles. Consequently, the normal resistivity would decrease as $\rho_N\sim a$. For reasonable values of particle sizes (60nm) this would lead to a critical current density of the order of $J_C\simeq 10^5$ A/cm$^2$ and therefore represent a very interesting system for various applications.

\section{Conclusion}

In summary, we discovered an efficient production path to making MgB$_2$ tapes through an initial formable step involving a cellulose based binder. The material exhibits a high transition temperature close to that of crystalline MgB$_2$. We showed that the binder does not alter any of the chemical properties and constitutes an optimal way to form a new compound, which is mechanically strong and formable. The main drawback is the very low critical current density. However, the current density could be increased substantially by using much smaller particle sizes, hence opening the door for many potential applications for high temperature superconductivity and possibly new quantum interference devices based on the intrinsic Josephson junctions.


\section*{Acknowledgment}

Samir Elouatik is gratefully acknowledged for his help in analyzing the results as well as support from NSERC, FQRNT, RQMP and INTRIQ.




\begin{thebibliography}{1}


\bibitem{1} J. Nagamatsu, N. Nakagawa, T. Muranaka, Y. Zenitani, J. Akimitsu, Nature 410 (2001) 63-64.
\bibitem{2} S. Jin, H. Mavoori, C. Bower, R. B. van Dover, Nature 411 (2001) 563-565.
\bibitem{3} V. P. S. Awana, R. Rawat, A. Gupta, M. Isobe, K. P. Singh, A. Vajpayee, H. Kishan, E. Takayama-Muromachi, A. V. Narlikar, Solid State Communications 139 (2006) 306-309.
\bibitem{4} Y. Bugoslavsky, G. K. Perkins, X. Qi, L. F. Cohen, A. D. Caplin, Nature 410 (2001) 563-565.
\bibitem{magnet} K. Vinod, R.G. Abhilash Kumar and U. Syamaprasad, Supercond. Sci. Technol. 20 (2007) R1–R13.
\bibitem{5} M. E. Jones, R. E. Marsh, J. Am. Chem. Soc. 76 (1954) 1434-1436.
\bibitem{PIT} S. Jin, H. Mavoori, C. Bower and R.B. van Dover, Nature 411 (2001), 563-565.
\bibitem{BCS} J. Bardeen, L. N. Cooper, and J. R. Schrieffer, Phys. Rev. 106, (1947) 162-164.
\bibitem{double} E.J. Choi, D. Roundy, H. Sun, M.L. Cohen, S.G. Loule, Nature 418, (2002) 758-760.
\bibitem{6} V. Suthar, A. Pratap, H. Raval, Bulletin of Materials Science 23 (2000) 215-219.
\newpage
\bibitem{7} C. S. Sundar, A. Bharathi, M. Premila, T. N. Sairam, S. Kalavathi, G. L. N. Reddy, V. S. Sastry, Y. Hariharan, T. S. Radhakrishnan, Condensed Matter (2001).
\bibitem{8} K. M. Subhedar, R. S. Hyam, V. Ganesan, S. H. Pawar, Physica C: Superconductivity 449 (2006) 73-79.
\bibitem{9} J. Kortus, I. I. Mazin, K. D. Belashchenko, V. P. Antropov, L. L. Boyer, Physical Review Letters 86 (2001) 4656.
\bibitem{Shigeta} I. Shigeta, T. Abiru, K. Abe, A. Nishida, Y. Matsumoto, Physica C: Superconductivity 392–396 (2003) 359–363.
\bibitem{Bean} C.P.
Bean, Rev. Mod. Phys. 36 (1964) 31.
\bibitem{Brandt} E.H. Brandt, Physica B 284-288 (2000) 743-744.
\bibitem{Xu} M. Xu, H. Kitazawa, Y. Takano, J. Ye, K. Nishida, H. Abe, A. Matsushita, N. Tsujii,
and G. Kido, Appl. Phys. Lett., 79 (2001) 2779-2781.
\bibitem{JJ} V. Ambegaokar and A. Baratoff, Phys. Rev. Lett. 10, (1963) 486-489.
\bibitem{JJN} J. R. Clem, B. Bumble, S. I. Raider, W. J. Gallagher, and Y. C. Shih, Phys. Rev. B 35, (1987) 6637-6642.






\end{thebibliography}
%

\end{document}